\documentclass[acmsmall,screen]{acmart}
\usepackage{graphicx}

\AtBeginDocument{%
  }

\begin{document}

\title{Evaluating More with Less: Black-Box Test Selection for AI Agents}

\author{Ali Asgari}
\affiliation{%
  \institution{Delft University of Technology}
  \city{Delft}
  \country{The Netherlands}
}
\email{a.asgari-2@tudelft.nl}

\author{Mitchell Olsthoorn}
\affiliation{%
  \institution{Delft University of Technology}
  \city{Delft}
  \country{The Netherlands}
}
\email{M.J.G.Olsthoorn@tudelft.nl;}

\author{Arie van Deursen}
\affiliation{%
  \institution{Delft University of Technology}
  \city{Delft}
  \country{The Netherlands}
}
\email{arie.vandeursen@tudelft.nl}

\author{Annibale Panichella}
\affiliation{%
  \institution{Delft University of Technology}
  \city{Delft}
  \country{The Netherlands}
}
\email{a.panichella@tudelft.nl}
\renewcommand{\shortauthors}{Asgari et al.}

\begin{abstract}
Agentic software systems are increasingly used in development workflows, where changes to the model, prompt, scaffold, tools, or execution environment require repeated testing. Selecting a subset of tests can reduce testing costs while providing quick estimates of agent performance or exposing failures early to support failure analysis. Although test case selection (TCS) has been studied for deep neural networks and LLM code models, prior work does not directly address agentic systems and often requires model information that is unavailable for proprietary agents.

This paper presents the first empirical study of black-box TCS for agentic software systems. We treat each benchmark task as a test case and adapt black-box regression-testing techniques by redefining their selection criteria over task descriptions, metadata, code fragments, and embedding representations. We also propose new strategies that estimate task difficulty using heuristic features or independent LLM-as-judge assessments. Our proposed and adapted strategies select tasks without executing the target agent or using past results.

We evaluate fourteen selection strategies, including random sampling, and an additional historical-data baseline on ten AI agent benchmarks, including SWE-bench Verified, with 12 to 101 agents per benchmark. We assess Success Rate estimation and early failure detection. No single strategy dominates both objectives. Adapted black-box regression-testing strategies are most effective for Success Rate estimation, with Category Partition obtaining the lowest median error on five of the ten benchmarks at the 20\% task budget. LLM-as-judge selection is more effective for early failure detection. For GPT 5.2 Codex on SWE-bench Verified, Category Partition executes 100 instead of 500 tasks and reduces execution cost from \$225 to \$47, with an absolute error in Success Rate estimation of 0.8 percentage points. At the same task budget, LLM-as-judge selection detects nearly twice as many failures as random sampling: 53.0\% of selected SWE-bench Verified tasks reveal failures, compared with 27.4\% for random sampling.
\end{abstract}


\maketitle

\section{Introduction}
\label{sec:intro}

AI agents increasingly write code, resolve programming issues, interact with the web, and perform a wide range of software engineering tasks. Their performance is typically evaluated using standard benchmarks that assess success in completing predefined tasks. As these agents evolve, modifications to the underlying model, prompt, orchestration, tools, or execution environment require continuous testing~\cite{asgari2026challenges}. However, executing a complete benchmark can be costly~\cite{hasan2026testing}: each task may require multiple calls to large language models (LLMs), tool interactions, and code execution. These costs increase with benchmark size. For example, SWE-bench Verified contains 500 tasks~\cite{jimenez2024swebench}, while SWE-Bench Pro contains 607~\cite{swebenchpro}. Selecting a subset of tasks could reduce these costs and provide faster feedback after changes by requiring fewer agent executions, provided that the selected tasks support accurate performance estimation or help reveal failures early.

A similar challenge has long existed in software testing. As test suites grow, executing every test after each software change becomes increasingly expensive~\cite{yoo2012regression}. Test case selection (TCS) addresses this issue by selecting a subset of tests~\cite{rothermel2001prioritizing,elbaum2002test}. More recently, TCS has been studied for deep neural networks (DNNs), primarily on image-classification tasks, and for LLM code models on classification tasks such as vulnerability detection, clone detection, and technical debt prediction~\cite{asgari2026test}. These studies consider two objectives: estimating overall model performance from a sample of inputs and selecting inputs that reveal failures early.

Applying TCS to agentic systems introduces new challenges. Many existing deep neural network (DNN) and LLM test-selection techniques depend on information from the system/model under test, such as confidence scores, uncertainty estimates, activations, coverage, or internal representations \cite{aghababaeyan2024deepgd, tian2018deeptest, abbasishahkoo2025metasel, asgari2026test}. However, such information is unavailable for closed-source models or black-box agents. Furthermore, agent evaluation typically involves generative tasks where the agent interacts with tools and environments to complete a task, rather than predicting a single class label.

In this paper, we formulate benchmark evaluation for agentic software systems as a test case selection problem. Each benchmark task is a \emph{test case}, each candidate agent is a \emph{system under test}, and an unsuccessful task is a \emph{failure}. 
Following prior test-selection studies~\cite{asgari2026test,guerriero2024deepsample,asgari2025adaptive}, we consider two objectives: estimating an agent's full-benchmark Success Rate from a subset of tasks and exposing failures early for failure analysis.
%
To the best of our knowledge, this is the first empirical study of black-box TCS for agentic software systems. We adapt seven black-box regression-testing techniques by redefining their selection criteria over task descriptions, metadata, code fragments, and embedding representations. These techniques include category partitioning, boundary-value selection, textual diversity, clustering, and adaptive random selection. We also propose six novel task-difficulty strategies: two use heuristic features, and four use independent LLM-as-judge assessments. These thirteen strategies require neither execution of the target agent nor recorded outcomes from the same or other agents.

Together with random sampling, they form the fourteen primary strategies evaluated in our study. We also include Historical Mid-Range, a baseline adapted from prior work on predicting agent leaderboard rankings~\cite{ndzomga2026efficient}. Unlike our strategies, this baseline uses recorded outcomes from other agents, excluding the target agent's outcomes. The study covers ten public AI agent benchmarks encompassing repository-level software engineering~\cite{jimenez2024swebench,yang2025swesmith,swebenchpro}, competitive programming~\cite{shi2024usaco}, terminal use~\cite{tbench2}, tool-using dialogue~\cite{yao2024taubench}, computational reproducibility~\cite{siegel2024corebench}, general assistant tasks~\cite{mialon2023gaia}, and web browsing~\cite{xue2025illusion}. 
For each strategy, we assess how accurately the selected set of tasks estimates each agent's full-benchmark Success Rate and how many failures it reveals at task budgets of 10\%, 20\%, and 50\%.


The results show that the two objectives benefit from different selection strategies. For Success Rate estimation, adapted black-box regression-testing strategies yield the strongest results, with Category Partition achieving the lowest median estimation error on five of the ten benchmarks. For early failure detection, LLM-as-judge difficulty-based selection proves more effective, prioritizing failing tasks earlier in the selected subset across the evaluated benchmarks. 

Our results also suggest that evaluation costs can be reduced in practice. For DeepSeek V3.2 on SWE-bench Multilingual, Category Partition evaluates 60 of 296 tasks, reducing execution cost by approximately 82\%, from \$114 to \$20, with an absolute error in Success Rate estimation of 0.8 percentage points. On SWE-Bench Pro, when testing GPT-5 within SWE-agent with a 250-turn limit, LLM-as-judge selection detects approximately 42\% more failures than random sampling at the same 20\% task budget: 91.7\% of selected tasks reveal failures, compared with 64.5\% for random sampling. These examples illustrate how task selection can reduce the cost of performance estimation and detect more failures within the same task budget.

This paper offers the following contributions:
\begin{itemize}
\item We formulate limited-budget evaluation of agentic software systems as a test case selection problem and, to the best of our knowledge, present the first empirical study of black-box TCS for agentic systems and generative agent tasks.
\item We adapt seven strategies from black-box regression testing and propose six task-difficulty strategies that select tasks without executing the target agent. We evaluate these strategies against random sampling and a historical-performance baseline.
\item We conduct an empirical study across ten public agent benchmarks, evaluating the strategies for Success Rate estimation and early failure detection at three task budgets using 12 to 101 agents per benchmark.
\item All code and results are released to support replication.
\end{itemize}


\section{Background and Related Work}

Regression testing aims to reduce the cost of executing large test suites after software changes~\cite{yoo2012regression}. Many regression testing techniques rely on white-box information, such as statement or branch coverage collected from previous executions. Black-box techniques, instead, use only information available from the test inputs and are therefore more practical when source code, instrumentation, or prior coverage data is unavailable. Henard et al.~\cite{henard2016comparing} showed that black-box techniques based on input characteristics and diversity can be competitive with coverage-based techniques. Related black-box testing methods include category partitioning~\cite{ostrand1988category}, where tests are derived from observable partitions of the input space, and Adaptive Random Testing (ART)~\cite{chen2010adaptive}, where each new test is selected to be far from the tests already selected according to a distance metric over the input domain. ART is motivated by the observation that failures often cluster in the input space, so diverse inputs can exercise different behaviors earlier.

Black-box regression testing has also used textual information to compare and select tests~\cite{feldt2016}. In information-retrieval-based regression testing, program changes are treated as queries and tests as documents to retrieve; models such as TF-IDF, LSI, and LDA are then used to rank tests by their textual similarity to changes~\cite{peng2020ir} or to other tests in the suite~\cite{aman2018}. More broadly, machine-learning-based regression testing has used test-case text to construct representations for unsupervised methods, including clustering and topic modeling~\cite{pan2022mltsp}.

Our setting differs from conventional regression testing because our test inputs are benchmark tasks rather than conventional executable test cases, such as JUnit tests. Each test is an agent task: a prompt together with the context, files, repository state, tools, and success criterion provided by the benchmark. Executing a test means invoking the agent on that task and observing whether it reaches the expected outcome. Thus, there is no executable test body comparable to a JUnit method; the executable procedure is the agent run, and the assertion is provided by the benchmark oracle or success criterion. We therefore re-interpret black-box regression testing criteria using task-side information, including task descriptions, benchmark metadata, code snippets, test names, and embedding representations.

\textbf{Test selection for deep models.} Test case selection has been studied for deep models to reduce labeling and execution costs while supporting performance estimation and early failure detection~\cite{ma2020testselection}. Selection criteria include uncertainty, surprise adequacy, coverage, sampling, and diversity~\cite{ma2020testselection,kim2019guiding,guerriero2024deepsample}. Their effectiveness depends on the objective: uncertainty-based criteria often favor failure discovery, whereas representative samples better support performance estimation. Many techniques require model-side information, such as probabilities, logits, or hidden-layer activations, which may be unavailable for proprietary agents or difficult to define consistently across systems combining LLMs, tools, memory, and execution environments.

Recent work has extended test case selection to LLMs. Clotho~\cite{yoon2026clotho} identifies inputs likely to reveal failures using LLM hidden states and a labeled reference set, before output generation. Asgari et al.~\cite{asgari2026test} re-examined DNN selection techniques on LLMs performing code-related classification tasks, such as clone detection, vulnerability detection, and technical-debt prediction. Several criteria in these studies rely on model-side information, such as predictive probabilities or hidden-layer representations. Our setting differs: we study agentic systems that solve generative, interactive, and often multi-step benchmark tasks. We focus on task-related criteria computed before executing the target agent, including task features, textual and embedding representations, and independent task-difficulty judgments, without requiring recorded agent outcomes.

\textbf{Predicting agent leaderboard rankings.}
Ndzomga studies reduced benchmark suites for predicting agent leaderboard rankings~\cite{ndzomga2026efficient}. The approach uses historical per-task pass rates from previously evaluated agents to identify tasks of intermediate difficulty, motivated by item response theory. Tasks that are solved by almost all agents or by almost none of them carry little information for ranking, whereas mid-difficulty tasks are expected to better discriminate among agents. The resulting reduced suite is evaluated by how well it preserves the ranking of agents that would be obtained from the complete benchmark.
Instead of finding a subset of tasks that can predict the overall ranking of agents, we search for a subset that facilitates predicting agent performance with a limited budget.  The selected prefix is assessed by how accurately it estimates the target agent's full-suite Success Rate and how many of its failures it exposes early. We adapt the historical mid-range idea as a baseline by computing each task's pass rate over all agents in the benchmark except the target agent, then prioritizing tasks whose pass rate is closest to 0.5. This leave-one-agent-out version keeps the original principle of selecting discriminative mid-difficulty tasks while avoiding the target agent's own outcomes when constructing the ordering.

\section{Test Case Selection for AI Agents}
\label{sec:methodology}

Following prior work on test selection for LLMs for code~\cite{asgari2026test}, we treat each benchmark task as a test case: an input on which the agent is executed and its success or failure is assessed. Tasks may include issue reports, programming problems, terminal instructions, web tasks, or tool-use scenarios.
Given a benchmark with tasks $T=\{t_1,\ldots,t_n\}$ and a budget of $k$ tasks, each strategy selects a subset $S\subseteq T$ with $|S|=k$. Selection follows the strategy's criterion until the budget is reached. For strategies that select tasks sequentially, the selected set contains the first $k$ tasks chosen. These sets support two objectives: estimating the agent's full-benchmark Success Rate and exposing failures early.


\subsection{Test Selection Techniques}
\label{sec:techniques}

We describe seven strategies adapted from black-box regression testing and six novel strategies that estimate task difficulty before executing the agent, together with random sampling as a baseline. Historical Mid-Range, an additional baseline requiring outcomes from other agents, is described in Section~\ref{sec:baselines}. Unless stated otherwise, ties are resolved by the original benchmark order: if tasks $t_i$ and $t_j$ receive the same score and $i<j$, then $t_i$ is selected first.

\subsubsection{Random sampling.}
Random is our feature-agnostic baseline~\cite{asgari2025adaptive,henard2016comparing}. For a budget of $k$ tasks, it samples $k$ tasks uniformly without replacement. This baseline tests whether task information, such as text, metadata, embeddings, or difficulty estimates, improves over chance. It is especially strong for Success Rate estimation because random samples are unbiased estimators of the full-benchmark Success Rate, but it does not target difficult or failure-prone tasks and is therefore less suited to early failure detection.

\subsubsection{Adapted Black-box Regression Techniques.}
The second family adapts black-box regression-testing ideas to agent benchmarks. In traditional black-box testing, selection uses observable properties of the test inputs rather than coverage or internal information from the system under test. We apply the same principle to agent tasks: the input is the benchmark task given to the agent, including its natural-language description, code snippets, metadata, test names, repository or task category, and other contextual fields. The following strategies reinterpret black-box selection criteria over these task-side properties.

\textbf{Task length} is a single-feature baseline. It computes the number of characters in the task text available before execution, including the instruction, problem statement, hints, contextual fields, and any code snippets that appear in them. The strategy selects longer tasks first, using length as a proxy for task complexity: longer tasks may contain more requirements, constraints, code fragments, or environmental details, and may require more context management, exploration, or tool use from the agent.

\begin{table}[t]
\centering
\caption{Selection criteria used by the evaluated strategies and  baselines}
\label{tab:criteria}
\setlength{\tabcolsep}{4pt}\scriptsize
\begin{tabular}{lll}
\toprule
\textbf{Family} & \textbf{Strategy} & \textbf{Selection criterion} \\
\midrule
\emph{Baselines} & Random & randomness only \\
& \textcolor{black}{Historical Mid-Range}~\cite{ndzomga2026efficient} & \textcolor{black}{distance of other agents' success rate from 50\%} \\
\midrule
\emph{Adapted Black-Box Regression} &
Task length & task-text length \\
& Boundary Value & distance from median over task-input features \\
& Category Partition & task categories, textual markers, length quartile \\
& TF-IDF Diversity (Original) & TF-IDF diversity over the original task text \\
& TF-IDF Diversity (Code) & TF-IDF diversity over task text plus extracted code \\
& Adaptive Random & CodeT5+ embeddings, UMAP-10, farthest-from-selected rule \\
& Round Robin & CodeT5+ $k$-means clusters, prose entropy, traceback \\
\midrule
\emph{Task-Side Difficulty Estimation} &
Difficulty Score & hand-designed difficulty score from task text and metadata \\
& Difficulty Score with Diversity & difficulty score plus cluster-based diversity \\
& LLM-as-Judge & one overall 1--4 difficulty judgment \\
& LLM-as-Judge with Diversity & one overall judgment plus cluster-based diversity \\
& Factor-Based LLM-as-Judge & several judged difficulty factors, $z$-scored and summed \\
& Factor-Based LLM-as-Judge with Diversity & factor-based judgment plus cluster-based diversity \\
\bottomrule
\end{tabular}

\vspace{2pt}
\end{table}


\textbf{Boundary Value} adapts boundary-value testing to agent tasks. Traditional boundary-value testing selects inputs near the edges of an input domain because such inputs are often more likely to expose faults. Since agent benchmarks do not provide explicit numeric input domains, we define boundaries over task-input features available before execution, such as task length, number and length of code spans, number of lines, token counts, prose entropy, and the presence of tracebacks or exception names. For each feature $f_j$, we compute the median $\widetilde{f_j}$ over the benchmark and assign each task a normalized distance from it: $d_j(t_i)=|f_j(t_i)-\widetilde{f_j}|/(\mathrm{MAD}(f_j)+\epsilon)$, where $\mathrm{MAD}$ is the median absolute deviation. The boundary score is $\mathrm{Boundary}(t_i)=\max_j d_j(t_i)$. Tasks are selected in decreasing score, so tasks that are unusually short, long, code-heavy, sparse, or otherwise extreme in at least one observable feature are selected earlier. The method does not use oracle information or agent outcomes.

\textbf{Category Partition} adapts the category-partition method~\cite{ostrand1988category}. The original method partitions the input domain using categories and choices derived from the specification. We apply the same principle to benchmark tasks by building categories from task information available before execution. When available, we use benchmark variables such as task type, domain, repository, language, or difficulty label; we also derive simple textual variables, such as whether the task contains a traceback, mentions an exception or warning, and its length quartile. For example, in SWE-bench-style repair tasks the repository can define a category variable, while in other benchmarks the task type or domain can play the same role. Each non-empty combination of variables defines a task category. The method sorts tasks within each category by decreasing length and then cycles over categories, selecting one task from each at a time. Early selected sets therefore cover different task types, domains, textual patterns, and length ranges.

\textbf{TF-IDF Diversity (Original and Code)} adapts diversity-based black-box regression testing to agent tasks~\cite{feldt2016,peng2020ir}. We evaluate two variants that differ only in how they construct the task document. \textit{TF-IDF Diversity (Original)} uses the original task text, including the instruction, problem statement, hints, context fields, and any existing code. \textit{TF-IDF Diversity (Code)} uses the same document but additionally appends extracted code blocks and inline code spans, increasing the contribution of code-related tokens to the representation without introducing new information.
Both variants use the same representation and selection procedure. We lowercase documents, tokenize them into word unigrams and bigrams, filter out English stop words, and represent them as TF-IDF vectors with sublinear term-frequency scaling. We do not apply stemming or lemmatization because identifiers, library names, and error names may carry useful information. Diversity between tasks $t_i$ and $t_j$ is computed using cosine distance,
$d(t_i,t_j)=1-\frac{\mathbf{x}_i \cdot \mathbf{x}_j}{\|\mathbf{x}_i\|\,\|\mathbf{x}_j\|}$,
where $\mathbf{x}_i$ and $\mathbf{x}_j$ are their TF-IDF vectors. Selection starts with one uniformly random task; each subsequent task maximizes its minimum distance to previously selected tasks. Neither variant uses agent outcomes or benchmark oracle information.

\textbf{Adaptive Random} adapts adaptive random testing~\cite{chen2010adaptive} to a task-embedding space. In ART, each new test is chosen to be far from the tests already selected. Here, task descriptions are embedded with CodeT5+ (\texttt{Salesforce/codet5p-110m-embedding})~\cite{wang2023codet5+}, reduced to 10 dimensions with UMAP using cosine distance, and normalized before selection. The method then applies the same farthest-from-selected rule as TF-IDF Diversity, but computes distances between task embeddings rather than TF-IDF vectors. Thus, early selected tasks cover different regions of the learned task representation.

\textbf{Round Robin} adapts clustering-based test selection and prioritization~\cite{pan2022mltsp}. The method clusters the same task embeddings used by Adaptive Random with $k$-means and then cycles over the non-empty clusters, selecting one task from each cluster before returning to the same cluster. Within each cluster, tasks are ordered by the sum of two standardized task-input features: prose-token entropy and the presence of a traceback. For a benchmark with $n$ tasks, we set $k=\max(4,\mathrm{round}(0.06n))$. This makes the number of clusters grow with benchmark size while retaining at least four clusters on small benchmarks~\cite{aloise2017strategies}. The value of $k$ is fixed before observing any agent outcome and is used only to obtain a coarse partition of the task space.

\subsubsection{Task-Side Difficulty Estimation}
The third family estimates task difficulty before executing the target agent. The rationale is that, under a limited testing budget, harder tasks may be more useful for exposing failures early. These strategies do not run the target agent and do not use agent outcomes. They use only task-side information available before execution, such as task descriptions, code fragments, metadata, or independent LLM-as-judge assessments. Each strategy constructs a benchmark-level task ordering, and the selected set at budget $k$ is the first $k$ tasks in that ordering. The same selected sets can therefore be reused for every agent evaluated on the benchmark.

\textbf{Difficulty Score} estimates five aspects of task difficulty from the task input. File, function, line-number, and traceback mentions reduce estimated localization difficulty. File-path mentions, code blocks, and text length approximate the expected change size; prose entropy, text length, and the proportion of code approximate reasoning difficulty. Counts of bullet points, numbered steps, and words expressing requirements (e.g., ``must'') or requested outputs (e.g., ``write'') approximate validation complexity. Reproduction-related words and code blocks reduce estimated reproducibility difficulty. The method standardizes features within each benchmark and combines them into five factor scores using predefined weights, documented in the replication package. It then standardizes the factor scores using $z$-scores and sums them with equal weight, selecting tasks with higher total scores first. 

\textbf{Difficulty Score with Diversity} uses the same difficulty score but combines it with diversity. The method first divides the tasks into four difficulty bands using the quartiles of the difficulty-score distribution within the benchmark. The bands are visited from hardest to easiest. Within each band, the method applies the Round Robin procedure over the task-embedding clusters, selecting one task from each non-empty cluster per cycle. 
Thus, the selected set favors difficult tasks while avoiding many near-duplicate tasks.

\textbf{LLM-as-Judge} uses an independent LLM to estimate the overall difficulty of each task before the target agent is executed. For each task, the judge is asked to assign one difficulty score on a four-point scale. The judge receives only task-side information available before target-agent execution, together with a benchmark-type description of what the task asks the agent to do and how success is determined. It does not receive target-agent outcomes, leaderboard information, benchmark labels, gold answers, oracle tests, or other outcome information. Tasks are selected in decreasing judged difficulty, with ties resolved by the original deterministic task order. 

\textbf{LLM-as-Judge with Diversity} uses the same one-dimensional LLM difficulty judgment but adds a diversity step. Tasks are first grouped by their judged difficulty score, from score 4 to score 1. The method visits these groups from hardest to easiest. Within each group, it applies cluster round-robin selection using task-embedding clusters, so that selected tasks remain diverse among tasks judged to have the same difficulty. This keeps the focus on hard tasks while reducing repeated selection of similar tasks. 

\textbf{Factor-Based LLM-as-Judge} asks an independent LLM to estimate difficulty along several predefined factors rather than giving only one overall score. For each task, the judge assigns a four-point score to each factor separately. The factors depend on the type of task. For coding and software-repair tasks, they include localization hardness, expected code-change size, reasoning hardness, validation-test complexity, and reproducibility hardness. For more general agent tasks, they include information gathering, action complexity, reasoning and planning, verification difficulty, and execution fragility. In both cases, the factor scores are standardized within the benchmark using $z$-scores and summed with equal weight. Tasks are selected in decreasing composite difficulty. 

\textbf{Factor-Based LLM-as-Judge with Diversity} combines the factor-based difficulty score produced by LLM-as-Judge with diversity. The method first computes the composite difficulty score from the standardized factor scores. It then divides tasks into four difficulty bands using the quartiles of this composite-score distribution and visits the bands from hardest to easiest. Within each band, it applies the Round Robin procedure over the task-embedding clusters, selecting one task from each non-empty cluster per cycle. The selected set therefore favors tasks judged difficult along multiple factors while spreading the budget across different task topics or regions. 

Section~\ref{sec:evaluation} provides additional details on the LLMs used as judges, the prompting procedure, the generated judgment scores, and the leakage-sensitivity analysis.

\section{Empirical Evaluation}
\label{sec:evaluation}

\textcolor{black}{This section describes the experimental protocol of our empirical study. We structure the study around two research questions, one for each evaluation objective established in the test-selection literature~\cite{hu2024test}: Success Rate estimation and early failure detection.}

\textcolor{black}{We investigate the following research questions:
\begin{itemize}
\item \textbf{RQ1:} How effective are selection strategies at estimating an agent's full-benchmark Success Rate under a limited task budget?
\item \textbf{RQ2:} How effective are selection strategies at detecting an agent's failures early under a limited task budget?
\end{itemize}}



\subsection{Benchmarks}
\label{sec:benchmarks}
We use ten public agent benchmarks covering issue resolution~\cite{jimenez2024swebench,yang2025swesmith,swebenchpro}, competitive programming~\cite{shi2024usaco}, terminal use~\cite{tbench2}, tool-using dialogue~\cite{yao2024taubench}, computational reproducibility~\cite{siegel2024corebench}, general assistant tasks~\cite{mialon2023gaia}, and web browsing~\cite{xue2025illusion}. We obtain published outcomes from the benchmark releases and the Holistic Agent Leaderboard (HAL)~\cite{stroebl2025hal}. 

Table~\ref{tab:benchmarks} summarizes the ten benchmarks, which contain 45 to 607 tasks. For each benchmark, published evaluation results are available for 12 to 101 agents and record whether each agent successfully completed each task. Here, an agent denotes an evaluated system configuration rather than a distinct underlying model: entries differ in their model, scaffold, tools, or execution limits. We retain the configurations associated with the published results without modifying or rerunning them. We use these outcomes to assess Success Rate estimation and failure detection on the tasks selected by each strategy, without rerunning the agents. None of the fourteen primary strategies uses agent outcomes to select tasks.

The benchmarks also differ in difficulty for the AI agents: the median failure rate across agents ranges from 27.2\% on SWE-bench Verified to 76.7\% on CORE-Bench Hard. Because benchmark sizes differ, we evaluate selection at 10\%, 20\%, and 50\% of the available tasks rather than using fixed task counts. SWE-bench Mini contains 50 tasks from SWE-bench Verified. We retain it as a separate setting because it includes a different set of evaluated agents and represents a smaller evaluation suite, while acknowledging that the two benchmarks overlap.

\begin{table}[!t]
\centering
\caption{The ten benchmark settings used in this study. Failure rate is the median per-agent failure rate within each benchmark.}
\label{tab:benchmarks}
\setlength{\tabcolsep}{5pt}\scriptsize
\begin{tabular}{llrrr}
\toprule
\textbf{Benchmark} & \textbf{Task} & \textbf{Agents} & \textbf{Tasks} & \textbf{Failure\ rate} \\
\midrule
SWE-bench Verified \cite{jimenez2024swebench}    & Issue resolution        & 13  & 500 & 0.272 \\
SWE-bench Multilingual \cite{yang2025swesmith}   & Issue resolution        & 13  & 296 & 0.328 \\
SWE-Bench Pro \cite{swebenchpro}                 & Issue resolution        & 12  & 607 & 0.731 \\
USACO \cite{shi2024usaco}                        & Competitive programming & 13  & 307 & 0.609 \\
TerminalBench 2.0 \cite{tbench2}                 & Terminal / CLI tasks    & 101 & 89  & 0.629 \\
TAU-bench Airline \cite{yao2024taubench}         & Tool-use dialogue       & 26  & 50  & 0.600 \\
CORE-Bench Hard \cite{siegel2024corebench}       & Research reproduction   & 38  & 45  & 0.767 \\
SWE-bench Mini \cite{stroebl2025hal}             & Issue resolution        & 28  & 50  & 0.720 \\
GAIA \cite{mialon2023gaia}                       & General assistant tasks & 32  & 164 & 0.515 \\
Online-Mind2Web \cite{xue2025illusion}           & Web navigation          & 22  & 240 & 0.637 \\
\bottomrule
\end{tabular}
\end{table}


\subsection{Baselines}
\label{sec:baselines}
\textcolor{black}{We compare our strategies with two baselines. Random sampling is a standard baseline in regression testing~\cite{rothermel2001prioritizing,elbaum2002test} and in test selection for deep neural networks~\cite{asgari2026test}. We use it to assess whether the other strategies estimate Success Rate more accurately or detect more failures within the same testing budget.}

\textcolor{black}{\textbf{Historical Mid-Range.}
Historical Mid-Range is an external baseline adapted from prior work~\cite{ndzomga2026efficient}. The original method selects benchmark tasks whose historical Success Rate across agents falls within a mid-difficulty range (30--70\%), with the goal of identifying tasks that are informative for distinguishing among agents. Since our setting instead requires selecting tasks under a limited evaluation budget for a single target agent, we adapt this idea into a complete task ordering.}

\textcolor{black}{For each target agent $a$, we compute the historical Success Rate of each task $t$ using the outcomes of all other agents on the same benchmark, $p_t^{(-a)} = \frac{1}{|A|-1}\sum_{r \in A,\;r\neq a} Y_{r,t}$, where $Y_{r,t}=1$ if agent $r$ successfully completes task $t$ and $0$ otherwise. Tasks are then ordered by $|p_t^{(-a)}-0.5|$ in ascending order, such that tasks with historical Success Rates closest to 50\% are selected first. This converts the original mid-range criterion into a full ordering from which the first $k$ tasks can be selected for any evaluation budget.}

\textcolor{black}{To avoid leakage, we exclude the target agent's own outcomes when constructing its ordering and use them only afterward to evaluate the selected tasks. Unlike our fourteen primary strategies, Historical Mid-Range therefore requires historical outcomes from other agents on the same benchmark. We consequently treat it as an external baseline rather than one of the fourteen primary strategies, while including it in our statistical comparisons.}

\subsection{Experimental Setup and Metrics}
We evaluate the strategies described in Section~\ref{sec:methodology} on the benchmarks introduced in Section~\ref{sec:benchmarks}. For each benchmark and target agent, each strategy generates an ordering of the benchmark tasks using only information available prior to evaluating the target agent. 

\paragraph{LLM judges.}
The main study uses Sonnet judgments. Newly collected judgments use \texttt{claude} \texttt{sonnet-5} with high reasoning effort, tools disabled, no web access, and no conversation history. To assess sensitivity to the judge model and possible benchmark contamination, we repeat the four judge strategies using \texttt{gpt-4o-mini-2024-07-18}, with temperature 0 and a maximum output length of 512 tokens. Its stated training cutoff predates the release of several evaluated benchmarks, allowing us to examine whether the results also hold with an older judge. Section~\ref{sec:discussion} discusses this analysis, and Section~\ref{sec:threats} addresses its limitations.
The five judged factors depend on task type: coding tasks use localization hardness, expected change size, reasoning hardness, validation complexity, and reproducibility hardness; other tasks use information gathering, action complexity, reasoning and planning, verification difficulty, and execution fragility. Scores are standardized per benchmark and summed equally. Newly collected responses must contain the required integer scores from 1 to 4; invalid responses are retried, and missing scores are not imputed. Prompts, available raw outputs, validated scores, settings, and selection scripts are included in our replication package.

\paragraph{Task budgets and repetitions.}
Following previous TCS studies on deep learning systems and LLM code models~\cite{guerriero2021operation,guerriero2024deepsample,asgari2025adaptive,asgari2026test}, we evaluate each strategy at budgets of 10\%, 20\%, and 50\% of benchmark tasks, rounded up to the nearest integer. These budgets represent small, moderate, and large partial evaluations, allowing us to assess whether performance patterns remain consistent as more tasks are evaluated. We run deterministic strategies once and repeat strategies involving random or seed-dependent steps 100 times with different seeds.

\paragraph{Outcome classification.}
We use published binary task outcomes where available. For fractional scores or success rates across repeated runs, a task counts as successful at 50\% or above. 

\paragraph{Success Rate estimation error (RQ1).}
Success Rate is the proportion of tasks that an agent completes successfully according to the benchmark's success criterion, such as passing the required tests for a software-repair task or completing the requested web interaction. For a target agent $a$ and budget $k$, let $\hat{R}_a(k)$ denote its Success Rate on the $k$ selected tasks and $R_a$ its Success Rate on the complete benchmark. The estimation error is defined as
\[
\mathrm{SccErr}_a(k)=\left|\hat{R}_a(k)-R_a\right|.
\]
Lower values indicate more accurate estimates. An error of zero means that the selected set and the complete benchmark yield the same Success Rate.

\paragraph{Failures found (RQ2).}
For a target agent $a$ and budget $k$, we count the selected tasks on which the agent does not satisfy the benchmark's success criterion. Higher values indicate that more failures are detected within the same task budget $k$.

\paragraph{Aggregation and statistical analysis.}
For randomized strategies, each target agent contributes the mean metric value across the 100 repetitions. For each benchmark, strategy, and budget, we report the median of the per-agent values and their interquartile range (IQR). The statistical tests use the per-agent values, not the benchmark-level medians.

For each metric and budget, we compare each technique family together with Random and Historical Mid-Range using the Friedman test at $\alpha=0.05$~\cite{Garcia:2009}. Strategies are the compared treatments, and benchmark--agent pairs from the ten benchmarks are the matched blocks. When Friedman identifies significant differences, we apply the Nemenyi post-hoc test to determine which strategies differ significantly~\cite{Garcia:2009}. Lower mean ranks indicate lower estimation error or more detected failures, depending on the objective. Two strategies differ significantly when the difference between their mean ranks exceeds the Nemenyi critical distance, which accounts for all pairwise comparisons within each family and its baselines. This combination of Friedman and post-hoc tests is widely used for comparing algorithms across multiple problems, including applications in search-based software engineering~\cite{devroey2023juge,panichella2021systematic}, benchmarking studies~\cite{japkowicz2011evaluating}, and a recent study on test selection for LLMs for code~\cite{asgari2026test}.

We evaluate all strategies at the 10\%, 20\%, and 50\% budgets.
Due to space limitations, Sections~\ref{sec:rq1} and~\ref{sec:rq2} present detailed results for the 20\% budget only. Complete results for all budgets are available in our replication package, and the effects of different budgets are discussed in Section~\ref{sec:discussion}.


\section{Results}

\subsection{RQ1: Success Rate Estimation}\label{sec:rq1}

\begin{table}[t]
\centering
\caption{Success Rate Estimation error (SccErr) at the 20\% evaluation budget across the ten benchmark settings.}
\label{tab:rq1}

\setlength{\tabcolsep}{5pt}\scriptsize
\resizebox{\linewidth}{!}{%
\begin{tabular}{lrrrrrrrrrr}
\toprule
\textbf{Strategy} & \textbf{SWE-V} & \textbf{USACO} & \textbf{T-Bench} & \textbf{TAU} & \textbf{CORE} & \textbf{Mini$^\ddagger$} & \textbf{Multi} & \textbf{Pro} & \textbf{GAIA} & \textbf{Mind2Web} \\
\midrule
\multicolumn{11}{l}{\emph{Baselines}} \\
Random & 0.033 (0.003) & 0.042 (0.003) & 0.077 (0.012) & 0.108 (0.023) & 0.093 (0.033) & 0.106 (0.055) & 0.042 (0.004) & 0.030 (0.005) & 0.059 (0.006) & 0.048 (0.004) \\
\textcolor{black}{Historical Mid-Range \cite{ndzomga2026efficient}} & \textcolor{black}{0.126 (0.142)} & \textcolor{black}{0.123 (0.161)} & \textcolor{black}{0.138 (0.159)} & \textcolor{black}{0.079 (0.045)} & \textcolor{black}{0.311 (0.262)} & \textcolor{black}{0.215 (0.326)} & \textcolor{black}{0.099 (0.087)} & \textcolor{black}{0.204 (0.096)} & \textcolor{black}{0.050 (0.066)} & \textcolor{black}{0.136 (0.119)} \\
\midrule
\multicolumn{11}{l}{\emph{Adapted Black-Box Regression}} \\
Task length & 0.082 (0.032) & 0.055 (0.034) & 0.044 (0.054) & 0.060 (0.120) & 0.222 (0.156) & 0.050 (0.065) & 0.093 (0.054) & 0.172 (0.066) & 0.058 (0.060) & 0.206 (0.087) \\
Boundary Value & 0.082 (0.032) & \textbf{0.024 (0.031)} & 0.044 (0.054) & \textbf{0.040 (0.105)} & 0.111 (0.083) & 0.050 (0.065) & 0.093 (0.054) & 0.119 (0.050) & 0.058 (0.060) & 0.146 (0.090) \\
Category Part. & \textbf{0.018 (0.018)} & 0.138 (0.037) & \textbf{0.041 (0.072)} & 0.130 (0.135) & \textbf{0.044 (0.089)} & 0.090 (0.125) & 0.033 (0.034) & 0.036 (0.039) & \textbf{0.047 (0.063)} & \textbf{0.017 (0.032)} \\
TF-IDF (Orig.) & 0.058 (0.020) & 0.069 (0.037) & 0.072 (0.020) & 0.099 (0.032) & 0.079 (0.031) & 0.084 (0.041) & 0.036 (0.007) & 0.051 (0.014) & 0.058 (0.031) & 0.083 (0.084) \\
TF-IDF (Code) & 0.045 (0.019) & 0.069 (0.037) & 0.071 (0.015) & 0.099 (0.032) & 0.079 (0.031) & 0.075 (0.032) & 0.035 (0.005) & 0.050 (0.023) & 0.058 (0.031) & 0.083 (0.084) \\
Adaptive Random & 0.021 (0.006) & 0.034 (0.008) & 0.069 (0.015) & 0.110 (0.024) & 0.094 (0.051) & 0.083 (0.067) & \textbf{0.032 (0.008)} & \textbf{0.029 (0.006)} & 0.062 (0.016) & 0.045 (0.024) \\
Round Robin & 0.140 (0.048) & 0.051 (0.051) & 0.056 (0.064) & 0.134 (0.103) & 0.183 (0.156) & 0.134 (0.119) & 0.100 (0.036) & 0.134 (0.032) & 0.061 (0.054) & 0.181 (0.058) \\
\midrule
\multicolumn{11}{l}{\emph{Task-Side Difficulty Estimation}} \\
Difficulty Score & 0.068 (0.038) & 0.055 (0.046) & 0.082 (0.078) & 0.110 (0.180) & 0.222 (0.150) & 0.060 (0.080) & 0.123 (0.048) & 0.125 (0.062) & 0.064 (0.084) & 0.185 (0.096) \\
Difficulty Score (Div.) & 0.118 (0.022) & 0.041 (0.046) & 0.059 (0.063) & 0.150 (0.115) & 0.222 (0.150) & \textbf{0.040 (0.080)} & 0.130 (0.024) & 0.158 (0.057) & 0.073 (0.086) & 0.179 (0.076) \\
LLM-as-Judge & 0.066 (0.012) & 0.253 (0.098) & 0.074 (0.079) & 0.080 (0.095) & 0.111 (0.150) & 0.060 (0.065) & 0.126 (0.048) & 0.215 (0.039) & 0.142 (0.105) & 0.196 (0.056) \\
LLM-as-Judge (Div.) & 0.186 (0.020) & 0.211 (0.080) & 0.091 (0.109) & 0.130 (0.135) & 0.211 (0.156) & 0.100 (0.100) & 0.127 (0.038) & 0.211 (0.060) & 0.127 (0.109) & 0.237 (0.058) \\
Factor-Based LLM-as-Judge & 0.258 (0.014) & 0.293 (0.042) & 0.180 (0.102) & 0.140 (0.075) & 0.156 (0.156) & 0.250 (0.285) & 0.137 (0.103) & 0.172 (0.053) & 0.131 (0.114) & 0.265 (0.067) \\
Factor-Based LLM-as-Judge (Div.) & 0.238 (0.032) & 0.281 (0.061) & 0.182 (0.108) & 0.167 (0.116) & 0.156 (0.158) & 0.279 (0.320) & 0.105 (0.068) & 0.170 (0.070) & 0.129 (0.089) & 0.207 (0.043) \\
\bottomrule
\end{tabular}%
}

\par\smallskip
{\scriptsize Values are median (IQR) across target agents, with each per-agent value averaged over 100 runs for randomized strategies. \textbf{Bold} indicates the best median. $^{\ddagger}$SWE-bench Mini is not independent of SWE-bench Verified (Section~\ref{sec:benchmarks}). GAIA uses the same 164 tasks for all strategies. \textit{(Div.)} abbreviates \textit{with Diversity}.}
\end{table}

\begin{figure}[t]
\centering
\includegraphics[width= 1\linewidth]{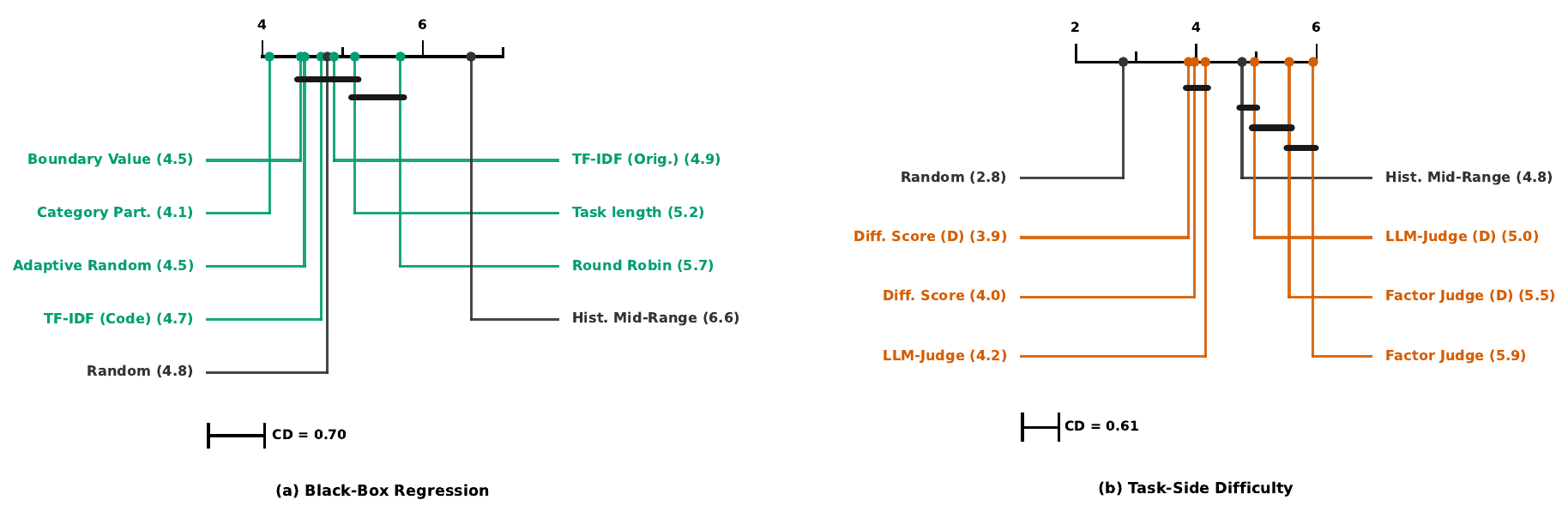}
\caption{Success Rate Estimation diagrams at the 20\% budget across ten benchmarks.}
\label{fig:cd-AccErr-k20}
\end{figure}

Table~\ref{tab:rq1} reports Success Rate estimation error at the 20\% task budget. Strategies adapted from black-box regression testing achieve the lowest median error on nine of the ten benchmarks: Category Partition on five, Adaptive Random on two, and Boundary Value on two. However, no individual strategy performs best across all benchmarks.

Category Partition produces substantial improvements over Random on several benchmarks. Its median error decreases from 3.3 to 1.8 percentage points on SWE-bench Verified, from 9.3 to 4.4 on CORE-Bench Hard, and from 4.8 to 1.7 on Online-Mind2Web. 
The two TF-IDF variants produce similar estimates. Giving additional weight to extracted code reduces median error on SWE-bench Verified from 5.8 to 4.5 percentage points, but the differences are smaller or absent on most other benchmarks. Diversity alone also does not guarantee accurate estimation. Round Robin has relatively high errors on several benchmarks, despite selecting tasks from different clusters.

\begin{figure*}[t]
\centering
\includegraphics[width=\textwidth]{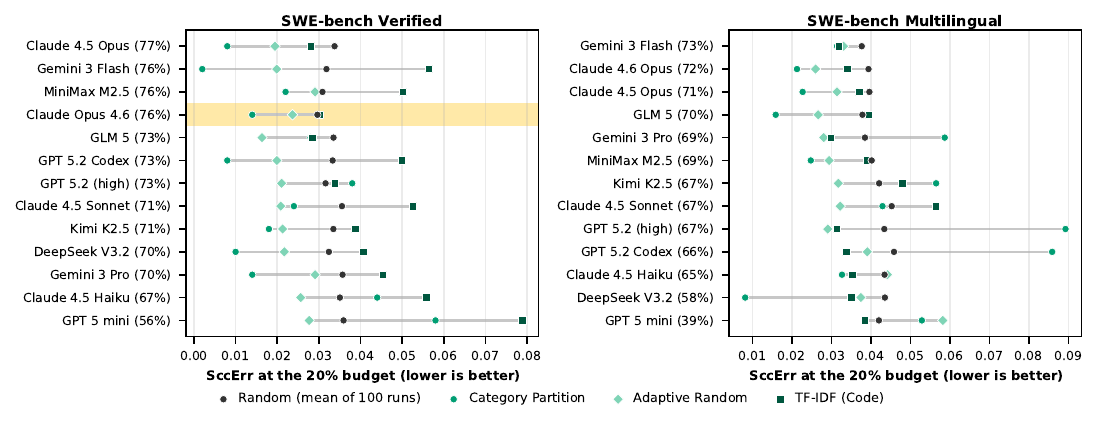}
\caption{\textcolor{black}{SccErr at the 20\% budget for each agent
Agents are sorted by true Success Rate (in brackets).}}
\label{fig:rq1-agents}
\end{figure*}

Difficulty-based selection generally produces larger estimation errors. On SWE-bench Verified, Factor-Based LLM-as-Judge has a median error of 25.8 percentage points, or 23.8 with diversity, compared with 3.3 for Random. Selecting difficult tasks may help expose failures but need not yield accurate full-benchmark estimates. Difficulty Score with Diversity is an exception, achieving the lowest median error on SWE-bench Mini. Historical Mid-Range also performs poorly on several benchmarks, showing that selecting tasks to distinguish among agents does not necessarily support accurate Success Rate estimation.
The benchmark-level results also reflect improvements for individual agents. For Claude Opus 4.6 on SWE-bench Verified, whose full-benchmark Success Rate is 75.6\%, Category Partition has an estimation error of 1.4 percentage points, compared with 3.0 for Random. Figure~\ref{fig:rq1-agents} shows the results for all 13 agents on SWE-bench Verified and SWE-bench Multilingual. Adaptive Random improves on Random for 13 and 11 agents, respectively, while Category Partition improves on it for 10 and 8 agents.

\textit{Statistical analysis.}
Figure~\ref{fig:cd-AccErr-k20} presents separate comparisons for the methods adapted from regression testing and the difficulty-based methods, each including Random and Historical Mid-Range as baselines. For the regression-testing methods (Figure~\ref{fig:cd-AccErr-k20}-a), the Friedman test identifies significant differences ($p$-value$<10^{-35}$). Category Partition obtains the best mean rank, and the Nemenyi post-hoc test ($CD=0.70$) shows that it significantly outperforms Random, although several other methods are not statistically distinguishable from Random.
For the difficulty-based methods (Figure~\ref{fig:cd-AccErr-k20}-b), the Friedman test also identifies significant differences ($p$-value$<10^{-77}$). The Nemenyi post-hoc test ($CD=0.61$) shows that Random significantly outperforms all six difficulty-based methods and Historical Mid-Range.

\noindent\fbox{\parbox{\dimexpr\linewidth-2\fboxsep-2\fboxrule\relax}{

\textbf{Finding RQ1: Success Rate Estimation.}
Adapted black-box strategies are the most effective for Success Rate estimation, achieving the lowest or joint-lowest median error on most benchmarks. Random itself remains a strong baseline.
}}

\subsection{RQ2: Early Failure Detection}\label{sec:rq2}

\begin{table}[t]
\centering
\caption{Failures detected at the 20\% evaluation budget across the ten benchmark settings.}
\label{tab:rq2}
\setlength{\tabcolsep}{5pt}\scriptsize
\resizebox{\linewidth}{!}{%
\begin{tabular}{lrrrrrrrrrr}
\toprule
\textbf{Strategy} & \textbf{SWE-V} & \textbf{USACO} & \textbf{T-Bench} & \textbf{TAU} & \textbf{CORE} & \textbf{Mini$^\ddagger$} & \textbf{Multi} & \textbf{Pro} & \textbf{GAIA} & \textbf{Mind2Web} \\
\midrule
\multicolumn{11}{l}{\emph{Baselines}} \\
Random & 27.4 (5.2) & 37.2 (12.0) & 11.2 (5.4) & 5.9 (2.6) & 7.0 (1.9) & 7.2 (3.8) & 19.0 (2.2) & 88.6 (15.2) & 17.1 (9.2) & 30.3 (5.1) \\
\textcolor{black}{Historical Mid-Range \cite{ndzomga2026efficient}} & \textcolor{black}{39.8 (19.6)} & \textcolor{black}{28.5 (21.5)} & \textcolor{black}{9.0 (10.0)} & \textcolor{black}{5.6 (2.5)} & \textcolor{black}{4.0 (4.5)} & \textcolor{black}{5.0 (7.0)} & \textcolor{black}{25.6 (6.2)} & \textcolor{black}{63.9 (26.2)} & \textcolor{black}{15.6 (11.5)} & \textcolor{black}{23.6 (12.3)} \\
\midrule
\multicolumn{11}{l}{\emph{Adapted Black-Box Regression}} \\
Task length & 34.0 (7.0) & 39.0 (11.0) & 10.0 (5.0) & 7.0 (2.8) & \textbf{9.0 (0.0)} & 8.0 (3.2) & 24.0 (5.0) & 106.5 (10.8) & 17.0 (8.2) & 41.0 (5.8) \\
Boundary Value & 34.0 (7.0) & 31.0 (12.0) & 10.0 (5.0) & 6.5 (3.0) & 8.0 (2.0) & 8.0 (3.2) & 24.0 (5.0) & 102.0 (8.8) & 17.0 (8.2) & 39.5 (5.8) \\
Category Part. & 29.0 (7.0) & 46.0 (13.0) & 10.0 (7.0) & \textbf{8.0 (2.0)} & 7.0 (2.0) & 5.0 (5.0) & 19.0 (8.0) & 92.5 (18.5) & 19.0 (8.0) & 31.0 (4.0) \\
TF-IDF (Orig.) & 22.7 (3.6) & 30.8 (11.7) & 10.6 (5.2) & 6.3 (2.7) & 6.6 (2.4) & 6.9 (3.4) & 18.2 (4.2) & 83.0 (15.3) & 16.5 (11.2) & 26.0 (4.6) \\
TF-IDF (Code) & 24.7 (3.9) & 30.8 (11.7) & 10.9 (6.0) & 6.3 (2.7) & 6.6 (2.4) & 7.0 (3.4) & 16.9 (3.2) & 82.1 (12.6) & 16.5 (11.2) & 26.0 (4.6) \\
Adaptive Random & 27.4 (3.1) & 36.9 (12.3) & 10.7 (5.4) & 6.1 (2.2) & 7.7 (1.4) & 6.9 (3.7) & 19.3 (2.1) & 87.3 (14.0) & 17.4 (9.9) & 31.7 (7.3) \\
Round Robin & 40.7 (8.5) & 42.2 (13.6) & 10.3 (6.0) & 7.6 (1.8) & 8.8 (0.5) & 9.0 (2.9) & 24.8 (1.9) & 105.6 (8.1) & 17.2 (7.6) & 39.6 (6.7) \\
\midrule
\multicolumn{11}{l}{\emph{Task-Side Difficulty Estimation}} \\
Difficulty Score & 34.0 (8.0) & 39.0 (9.0) & 10.0 (5.0) & 8.0 (3.0) & 9.0 (0.0) & 7.0 (3.2) & \textbf{27.0 (5.0)} & 104.0 (8.8) & 17.0 (8.0) & 40.0 (6.0) \\
Difficulty Score (Div.) & 39.0 (5.0) & 36.0 (10.0) & 10.0 (6.0) & 8.0 (2.8) & 9.0 (0.0) & 8.0 (3.0) & 27.0 (4.0) & 106.5 (9.0) & 16.0 (6.5) & 41.0 (6.8) \\
LLM-as-Judge & 34.0 (4.0) & 51.0 (7.0) & 12.0 (7.0) & 7.0 (2.8) & 6.0 (2.0) & 7.5 (3.2) & 26.0 (3.0) & \textbf{114.5 (7.0)} & 22.0 (7.2) & 40.0 (4.8) \\
LLM-as-Judge (Div.) & 46.0 (7.0) & 52.0 (7.0) & 12.0 (9.0) & 8.0 (3.0) & 9.0 (0.0) & 8.0 (3.2) & 26.0 (4.0) & 113.5 (7.5) & \textbf{23.0 (9.5)} & 41.5 (6.0) \\
Factor-Based LLM-as-Judge & \textbf{53.0 (6.0)} & 53.0 (8.0) & \textbf{15.0 (6.0)} & 8.0 (2.0) & 8.0 (1.0) & \textbf{10.0 (1.0)} & 26.0 (8.0) & 109.0 (8.2) & 23.0 (12.2) & \textbf{44.0 (3.0)} \\
Factor-Based LLM-as-Judge (Div.) & 51.1 (7.5) & \textbf{53.2 (5.7)} & 14.4 (4.9) & 7.6 (1.8) & 8.4 (1.0) & 10.0 (0.5) & 24.7 (5.7) & 109.2 (8.0) & 22.3 (10.5) & 41.7 (5.2) \\
\midrule
Max findable at 20\% budget & 100 & 62 & 18 & 10 & 9 & 10 & 60 & 122 & 33 & 48 \\
\bottomrule
\end{tabular}%
}

\par\smallskip
{\scriptsize Values are median (IQR) across target agents, with each per-agent value averaged over 100 runs for randomized strategies. \textbf{Bold} indicates the best median. $^{\ddagger}$SWE-bench Mini is not independent of SWE-bench Verified (Section~\ref{sec:benchmarks}). \textit{(Div.)} = \textit{Diversity}.}
\end{table}

\begin{figure}[t]
\centering
\includegraphics[width= 1\linewidth]{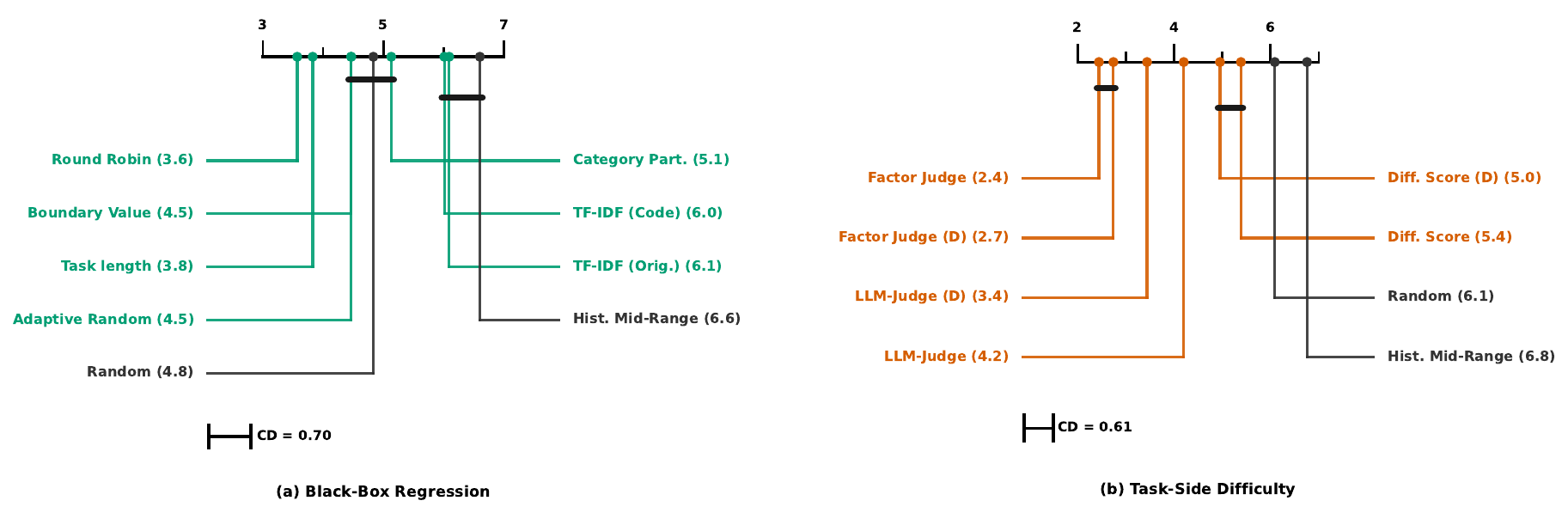}
\caption{Failure-detection at the 20\% budget across ten benchmarks: Lower mean rank is better.}
\label{fig:cd-failures-k20}
\end{figure}

Table~\ref{tab:rq2} reports the median number of failures detected at the 20\% task budget. Unlike Success Rate estimation, early failure detection benefits from difficulty-based selection. Factor-Based LLM-as-Judge achieves the highest or joint-highest median on six benchmarks, while its diversity variant achieves the highest median on USACO. LLM-as-Judge performs best on SWE-Bench Pro, and its diversity variant shares the highest median on GAIA. The strongest variant nevertheless depends on the benchmark: judging several difficulty factors is not always more effective than asking for one overall difficulty score.

The methods adapted from regression testing show more variable results. Round Robin obtains the best mean rank within this family, while Task length and Boundary Value are competitive on several benchmarks. In contrast, both TF-IDF variants generally detect fewer failures than Random. Difficulty-based selection can provide substantial improvements: on SWE-bench Verified, Factor-Based LLM-as-Judge detects a median of 53.0 failures, compared with 27.4 for Random, nearly twice as many within the same task budget. On Online-Mind2Web, it detects a median of 44.0 failures, compared with 30.3 for Random.

\begin{figure*}[t]
\centering
\includegraphics[width=\textwidth]{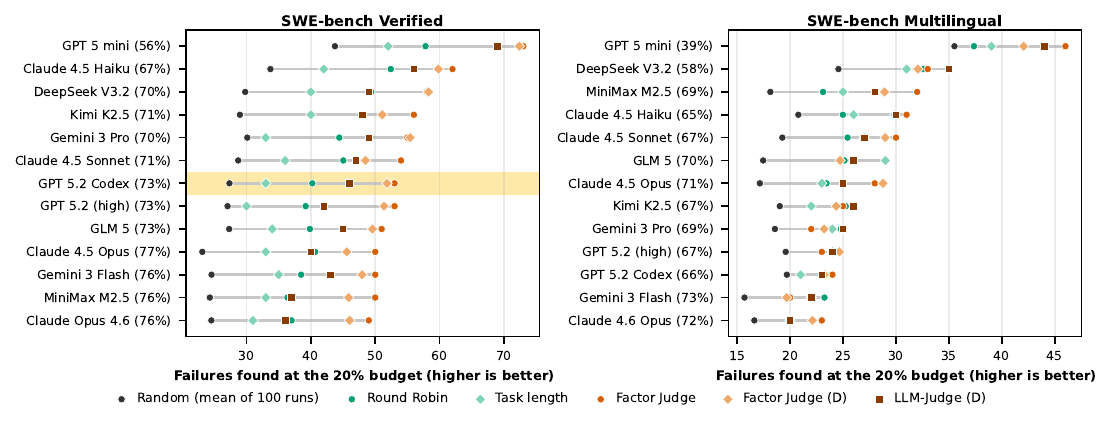}
\caption{\textcolor{black}{Failures found at the 20\% budget for each agent. 
Agents are sorted by the most failures found. Also see the true Success Rate (in brackets).}}
\label{fig:rq2-agents}
\end{figure*}

\textcolor{black}{As an example at the agent level, for GPT 5.2 Codex on SWE-bench Verified, Factor-Based LLM-as-Judge finds 53.0 failures at the 20\% budget, compared with 27.4 for Random and 40.2 for Round Robin, the best-ranked black-box strategy. The agent has 136 failures in the full benchmark, so evaluating 20\% of the tasks reveals about 39\% of them. Figure~\ref{fig:rq2-agents} shows that Factor-Based LLM-as-Judge finds more failures than Random for all 13 agents on both benchmarks and more than Round Robin for all 13 agents on SWE-bench Verified and 8 on SWE-bench Multilingual.}

\textit{Statistical analysis.}
%
%
Figure~\ref{fig:cd-failures-k20} presents separate comparisons for the methods adapted from regression testing and the difficulty-based methods, each including Random and Historical Mid-Range as baselines. Both comparisons use 298 benchmark--agent blocks. For the regression-testing methods (Figure~\ref{fig:cd-failures-k20}-a), the Friedman test identifies significant differences ($p$-value$<10^{-72}$). The Nemenyi post-hoc test ($CD=0.70$) shows that Round Robin and Task length significantly outperform Random. Round Robin obtains the best mean rank, followed by Task length, while Historical Mid-Range has the worst mean rank.
For the difficulty-based methods (Figure~\ref{fig:cd-failures-k20}-b), the Friedman test also identifies significant differences ($p$-value$<10^{-190}$). The Nemenyi post-hoc test ($CD=0.61$) shows that all six difficulty-based methods significantly outperform both baselines. Factor-Based LLM-as-Judge and its diversity variant obtain the best mean ranks and are not statistically distinguishable from each other. These results contrast with RQ1: difficulty-based selection is effective for exposing failures, but not for estimating full-benchmark Success Rate.

\noindent\fbox{\parbox{\dimexpr\linewidth-2\fboxsep-2\fboxrule\relax}{
\textbf{Finding RQ2: Early Failure Detection.}
Task-Side Difficulty Estimation strategies are the most effective for early failure detection, consistently outperforming Random across the benchmarks. And especially LLM-judge based strategies achieve the strongest results.
}}

\section{Discussion}
\label{sec:discussion}

\begin{table*}[t]
\centering
\small
\setlength{\tabcolsep}{2.5pt}
\caption{Cost of the strategies (median over 13 agents).}
\label{tab:eval-cost}
\resizebox{\textwidth}{!}{%
\begin{tabular}{@{}l rrr@{\hspace{6pt}}rrr@{\hspace{6pt}}rrr@{\hspace{12pt}}rrr@{\hspace{6pt}}rrr@{\hspace{6pt}}rrr@{}}
\toprule
 & \multicolumn{9}{c}{\textbf{SWE-bench Verified} \textbf{\large(whole suite \$225)}} & \multicolumn{9}{c}{\textbf{SWE-bench Multilingual} \textbf{\large(whole suite \$191)}} \\
\cmidrule(lr){2-10}\cmidrule(lr){11-19}
 & \multicolumn{3}{c}{10\%} & \multicolumn{3}{c}{20\%} & \multicolumn{3}{c}{50\%} & \multicolumn{3}{c}{10\%} & \multicolumn{3}{c}{20\%} & \multicolumn{3}{c}{50\%} \\
\cmidrule(lr){2-4}\cmidrule(lr){5-7}\cmidrule(lr){8-10}\cmidrule(lr){11-13}\cmidrule(lr){14-16}\cmidrule(lr){17-19}
Strategy & \$ & FD & SccErr& \$ & FD & SccErr& \$ & FD & SccErr& \$ & FD & SccErr& \$ & FD & SccErr& \$ & FD & SccErr\\
\midrule
Random & \textbf{23} & 14.1 & 0.050 & \textbf{45} & 27.4 & 0.033 & \textbf{113} & 68.6 & 0.016 & \textbf{19} & 9.7 & 0.061 & \textbf{37} & 19.0 & 0.042 & \textbf{95} & 48.8 & 0.021 \\
\textcolor{black}{Historical Mid-Range \cite{ndzomga2026efficient}} & \textcolor{black}{\textbf{34}} & \textcolor{black}{\underline{22.1}} & \textcolor{black}{0.163} & \textcolor{black}{\textbf{63}} & \textcolor{black}{\underline{39.8}} & \textcolor{black}{0.126} & \textcolor{black}{\textbf{133}} & \textcolor{black}{\underline{77.5}} & \textcolor{black}{0.038} & \textcolor{black}{\textbf{21}} & \textcolor{black}{\underline{13.7}} & \textcolor{black}{0.119} & \textcolor{black}{\textbf{46}} & \textcolor{black}{\underline{25.6}} & \textcolor{black}{0.099} & \textcolor{black}{\textbf{110}} & \textcolor{black}{\underline{53.9}} & \textcolor{black}{0.035} \\
\multicolumn{19}{@{}l}{\emph{Adapted Black-Box Regression}} \\
\quad TF-IDF (Orig.) & \textbf{19} & 9.7 & 0.089 & \textbf{41} & 22.7 & 0.058 & \textbf{108} & 65.8 & 0.021 & \textbf{20} & \underline{9.9} & \underline{0.057} & \textbf{38} & 18.2 & \underline{0.036} & \textbf{93} & 48.1 & \underline{\textbf{0.014}} \\
\quad TF-IDF (Code) & \textbf{21} & 10.3 & 0.072 & \textbf{44} & 24.7 & 0.045 & \textbf{107} & 65.3 & \underline{0.016} & \textbf{19} & 9.5 & \underline{0.055} & \textbf{37} & 16.9 & \underline{0.035} & \textbf{91} & 43.7 & 0.022 \\
\quad Adaptive Random & \textbf{22} & 13.6 & \underline{0.035} & \textbf{46} & \underline{27.4} & \underline{0.021} & \textbf{109} & 65.1 & \underline{\textbf{0.013}} & \textbf{21} & \underline{10.4} & \underline{0.052} & \textbf{35} & \underline{19.3} & \underline{\textbf{0.032}} & \textbf{89} & 43.8 & 0.028 \\
\quad Category Partition & \textbf{23} & 13.0 & \underline{\textbf{0.032}} & \textbf{47} & \underline{29.0} & \underline{\textbf{0.018}} & \textbf{118} & \underline{79.0} & 0.028 & \textbf{16} & 9.0 & \underline{0.038} & \textbf{31} & 19.0 & \underline{0.033} & \textbf{81} & 42.0 & 0.034 \\
\quad Round Robin & \textbf{30} & \underline{21.6} & 0.160 & \textbf{57} & \underline{40.7} & 0.140 & \textbf{132} & \underline{87.8} & 0.079 & \textbf{21} & \underline{11.5} & 0.079 & \textbf{46} & \underline{24.8} & 0.100 & \textbf{103} & \underline{57.7} & 0.073 \\
\quad Task length & \textbf{30} & \underline{19.0} & 0.118 & \textbf{59} & \underline{34.0} & 0.082 & \textbf{133} & \underline{86.0} & 0.072 & \textbf{19} & \underline{11.0} & \underline{\textbf{0.036}} & \textbf{42} & \underline{24.0} & 0.093 & \textbf{105} & \underline{56.0} & 0.061 \\
\quad Boundary Value & \textbf{30} & \underline{19.0} & 0.118 & \textbf{59} & \underline{34.0} & 0.082 & \textbf{123} & \underline{72.0} & 0.020 & \textbf{19} & \underline{11.0} & \underline{\textbf{0.036}} & \textbf{42} & \underline{24.0} & 0.093 & \textbf{104} & \underline{52.0} & 0.024 \\
\multicolumn{19}{@{}l}{\emph{Task-Side Difficulty Estimation}} \\
\quad Difficulty Score & \textbf{31} & \underline{20.0} & 0.122 & \textbf{61} & \underline{34.0} & 0.068 & \textbf{125} & \underline{74.0} & 0.026 & \textbf{21} & \underline{13.0} & 0.093 & \textbf{40} & \underline{\textbf{27.0}} & 0.123 & \textbf{95} & \underline{57.0} & 0.071 \\
\quad Difficulty Score (Div.) & \textbf{33} & \underline{22.0} & 0.168 & \textbf{63} & \underline{39.0} & 0.118 & \textbf{125} & \underline{74.0} & 0.026 & \textbf{20} & \underline{13.0} & 0.112 & \textbf{43} & \underline{\textbf{27.0}} & 0.130 & \textbf{95} & \underline{57.0} & 0.071 \\
\quad LLM-as-Judge & \textbf{33} & \underline{23.0} & 0.176 & \textbf{64} & \underline{34.0} & 0.066 & \textbf{159} & \underline{\textbf{95.0}} & 0.092 & \textbf{33} & \underline{14.0} & 0.150 & \textbf{56} & \underline{26.0} & 0.126 & \textbf{117} & \underline{52.0} & 0.051 \\
\quad LLM-as-Judge (Div.) & \textbf{41} & \underline{27.0} & 0.274 & \textbf{73} & \underline{46.0} & 0.186 & \textbf{159} & \underline{92.0} & 0.088 & \textbf{34} & \underline{14.0} & 0.152 & \textbf{54} & \underline{26.0} & 0.127 & \textbf{119} & \underline{57.0} & 0.064 \\
\quad Factor-Based LLM-as-Judge (Div.) & \textbf{39} & \underline{27.3} & 0.267 & \textbf{77} & \underline{51.1} & 0.238 & \textbf{153} & \underline{94.0} & 0.112 & \textbf{27} & \underline{14.3} & 0.137 & \textbf{55} & \underline{24.7} & 0.105 & \textbf{124} & \underline{\textbf{58.0}} & 0.065 \\
\quad Factor-Based LLM-as-Judge & \textbf{42} & \underline{\textbf{29.0}} & 0.308 & \textbf{80} & \underline{\textbf{53.0}} & 0.258 & \textbf{153} & \underline{94.0} & 0.112 & \textbf{33} & \underline{\textbf{16.0}} & 0.216 & \textbf{61} & \underline{26.0} & 0.137 & \textbf{124} & \underline{58.0} & 0.071 \\
\bottomrule
\end{tabular}}
\par\smallskip
{\raggedright\footnotesize FD is median failures found (higher is better); SccErr is Success Rate Estimation error (lower is better). For FD and SccErr, \textbf{bold} marks the best value, and \underline{underline} marks any strategy value better than \textbf{Random}. All costs (\$) are shown in \textbf{bold}. \textit{(Div.)}= \textit{Diversity}.}
\end{table*}

\begin{table}[t]
\centering
\small
\caption{Judging cost by benchmark (USD).}
\label{tab:judge-cost}
\resizebox{\linewidth}{!}{%
\begin{tabular}{lrrrrrrrrrrr}
\toprule
 & SWE-V & USACO & T-Bench & TAU & CORE & Mini & Multi & Pro & GAIA & Mind2Web & Total \\
\emph{Released} & \emph{Aug 2024} & \emph{Apr 2024} & \emph{Nov 2025} & \emph{Jun 2024} & \emph{Sep 2024} & \emph{Oct 2025} & \emph{Apr 2025} & \emph{Sep 2025} & \emph{Nov 2023} & \emph{Apr 2025} & \\
\midrule
Sonnet 5 (Jun 2026) & \$8.088 & \$10.288 & \$1.613 & \$0.811 & \$0.741 & \$0.000 & \$6.235 & \$12.101 & \$3.028 & \$2.592 & \textbf{\$45.497} \\
GPT-4o mini (Cutoff: Oct 2023) & \$0.205 & \$0.114 & \$0.025 & \$0.014 & \$0.013 & \$0.000 & \$0.149 & \$0.304 & \$0.046 & \$0.064 & \textbf{\$0.934} \\
\bottomrule
\end{tabular}}
\par\smallskip
{\raggedright\footnotesize Computed from recorded token usage. Mini reuses Verified's judgments at no extra cost.\par}
\end{table}


\textbf{Extending test selection to agentic and generative tasks}.
Test selection under limited budgets has traditionally focused on predictive neural models, especially deep neural networks and vision classification tasks \cite{guerriero2024deepsample, abbasishahkoo2025metasel, kim2019guiding, pietrantuono2018probabilistic}, emphasizing performance estimation and early failure detection as key evaluation objectives\cite{aghababaeyan2024deepgd, guerriero2021operation, tian2018deeptest, zhou2020cost}. A previous study extended these objectives to large language models (LLMs) for code, addressing classification-based software engineering tasks such as vulnerability detection, clone detection, and technical debt prediction~\cite{asgari2026test}. The findings indicated that representation-based approaches were more effective for Success Rate Estimation, whereas uncertainty-based approaches were more effective for early failure detection.

This study advances the investigation of these two objectives by focusing on software engineering agents and generative tasks. Unlike classification settings, agent benchmarks assess whether an agent can complete a task through a sequence of actions and generated outputs. The predictive probabilities and internal model representations utilized by traditional deep neural network (DNN) test-selection techniques are not directly available or comparable across agents. Consequently, this study's strategies rely on information about the benchmark tasks themselves, including textual properties, embeddings, task categories, and difficulty estimates.

This is the first study to investigate both Success Rate estimation and early failure detection through task selection for software engineering agents and generative agent benchmarks. Across ten benchmarks, our results show that both objectives can be supported under limited task budgets, but they favor different categories of selection strategies.

\textbf{Success rate estimation versus failure detection}.
The results reveal a clear distinction between the two evaluation objectives. Strategies that perform well for Success Rate estimation generally differ from those effective for early failure detection. For Success Rate estimation, black-box strategies such as Category Partition, Adaptive Random, and Random perform well, as they do not explicitly prioritize tasks based on estimated difficulty. In contrast, early failure detection benefits from difficulty-based selection, particularly the LLM-judge strategies.
This distinction aligns with the differing goals of the two objectives. Estimating the full-benchmark Success Rate requires the evaluated tasks to approximate the complete benchmark effectively. In contrast, prioritizing difficult tasks intentionally places those with a higher estimated failure risk earlier in the ordering. This helps reveal failures earlier, but the resulting prefix can be less suitable for estimating performance over the complete benchmark. Conversely, a more representative ordering can provide an accurate estimate of the overall Success Rate without necessarily prioritizing failure-prone tasks.

\begin{figure*}[t]
\centering
\includegraphics[width=\textwidth]{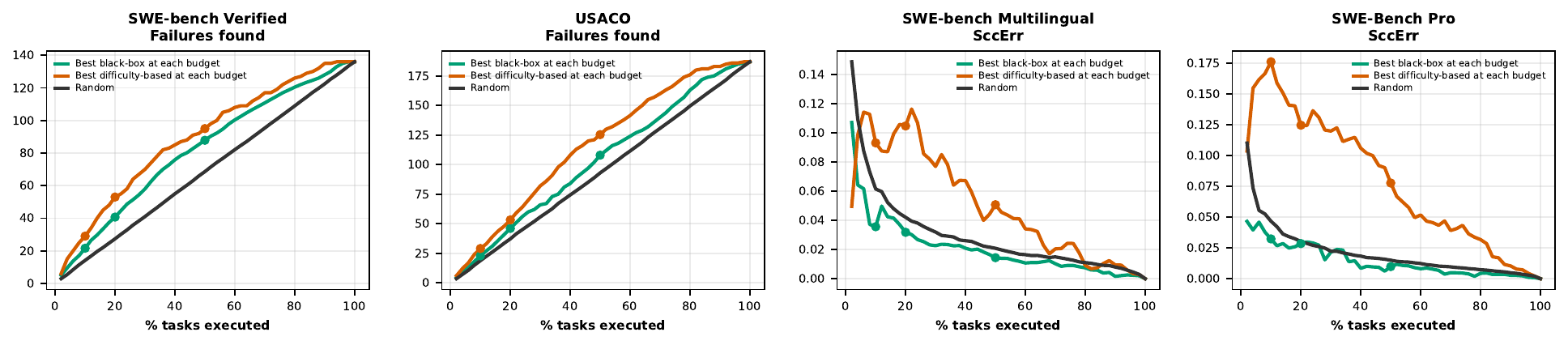}
\caption{\textcolor{black}{Failures found and SccErr vs.\ the fraction of tasks executed, on four benchmarks. Each figure shows Random and, at each budget, the best black-box and best difficulty-based strategy.}}
\label{fig:budget-growth}
\end{figure*}

\textcolor{black}{\textbf{Results across evaluation budgets.}
Figure~\ref{fig:budget-growth} shows how Failures Found and SccErr change as the evaluation budget increases from a small fraction of the benchmark to the full benchmark. We show two benchmarks for each objective: SWE-bench Verified and USACO for failure detection, and SWE-bench Multilingual and SWE-Bench Pro for SccErr. We compare Random with the best black-box strategy and the best difficulty-based strategy at each budget so that the best strategy can change along the curve; the dots mark the 10\%, 20\%, and 50\% budgets. Because we choose the best strategy at each budget, both family curves represent their best performance. The pattern observed at the 10\%, 20\%, and 50\% budgets holds across the full budget range on these benchmarks. For failure detection, the best difficulty-based strategy finds more failures than the best black-box strategy and Random at almost every budget. For SccErr, the best black-box strategy has the lowest error, while the best difficulty-based strategy rarely improves on Random. As the budget approaches 100\%, all strategies converge.
}

\begin{figure}[t]
\centering
 \includegraphics[width= 0.9\textwidth]{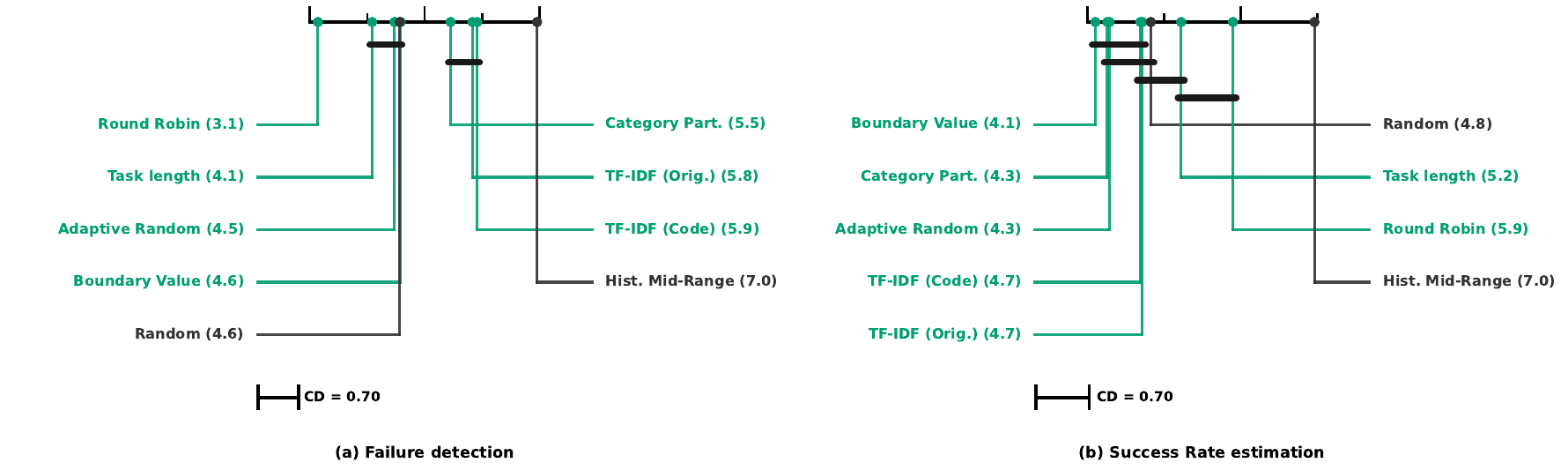}
\caption{Strategies adapted from black-box regression testing compared with the baselines across all benchmarks and task budgets.}
\label{fig:cd-overall-f2}
\end{figure}

\begin{figure}[t]
\centering
 \includegraphics[width= 0.9\linewidth]{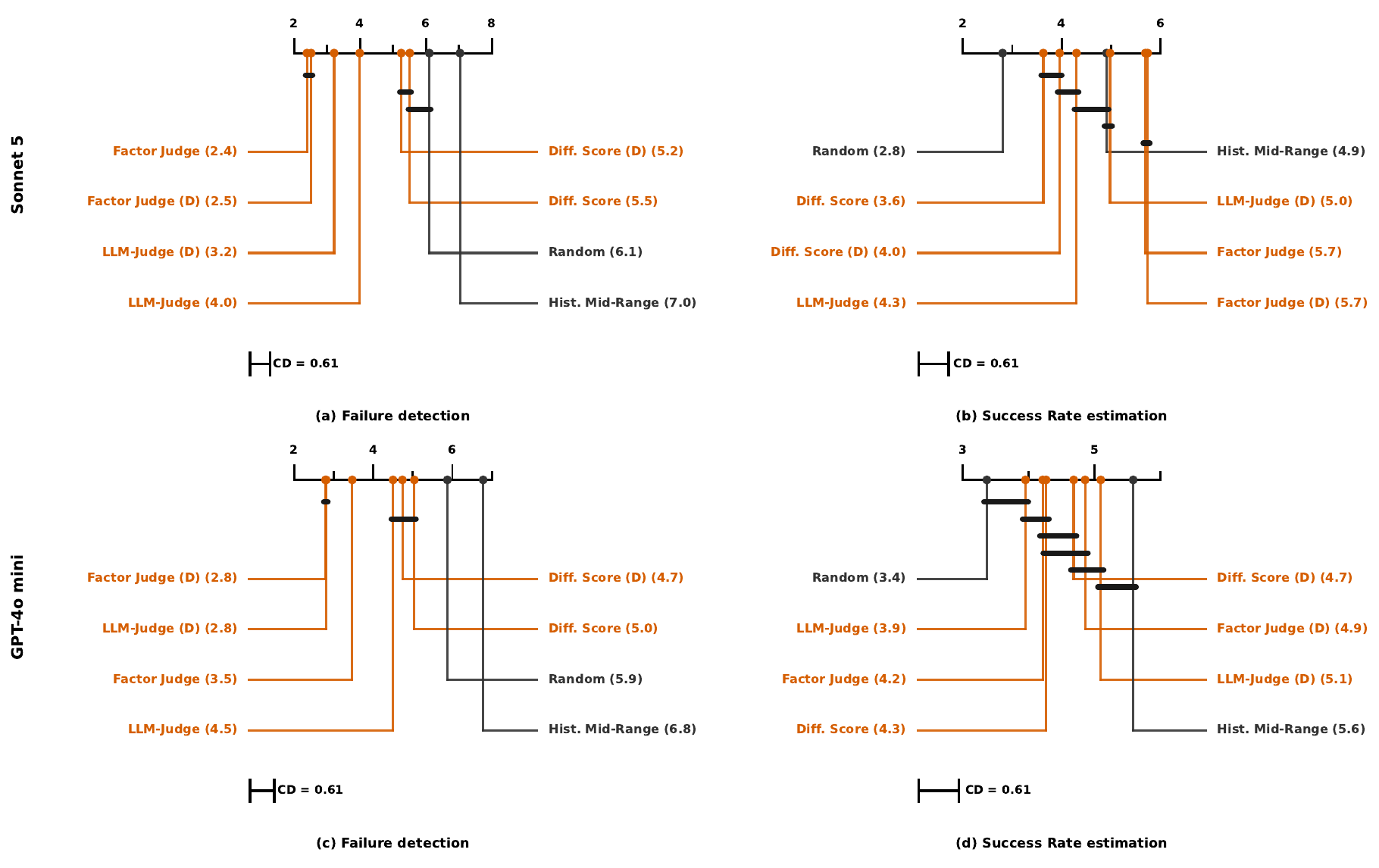}
\caption{Task-difficulty strategies compared with the baselines across all benchmarks and task budgets.}
\label{fig:cd-overall-f3}
\end{figure}

The statistical comparisons based on performance averaged across the 10\%, 20\%, and 50\% budgets show a similar overall ranking pattern (Figure~\ref{fig:cd-overall-f2} and Figure~\ref{fig:cd-overall-f3}). Category Partition, Boundary Value, and Adaptive Random obtain the best mean ranks for Success Rate estimation among the regression-testing methods and baselines. Factor-Based LLM-as-Judge variants remain best-ranked for failure detection, while Random remains best-ranked for Success Rate estimation among the difficulty-based methods and baselines.

\textbf{Impact on cost savings}. 
Reducing the number of evaluated tasks can substantially decrease agent evaluation costs. Table~\ref{tab:eval-cost} illustrates this using SWE-bench Verified and SWE-bench Multilingual. At the 20\% task budget, the evaluated strategies cost between \$41 and \$80 on SWE-bench Verified, compared with approximately \$225 for the complete benchmark. On SWE-bench Multilingual, the corresponding costs range from \$31 to \$61, compared with approximately \$191 for the complete benchmark.

However, task budget and monetary cost are not equivalent. Different strategies may execute the same number of tasks at a given budget, but the selected tasks can have varying execution costs. For example, at the 20\% budget on SWE-bench Verified, Category Partition costs \$47, while Factor-Based LLM-as-Judge costs \$80. Factor-Based LLM-as-Judge detects substantially more failures, with a median of 53.0 compared to 29.0 for Category Partition, but it also executes a more expensive set of tasks. This demonstrates that selection influences not only which failures are observed but also the monetary cost of the evaluation.

For Success Rate estimation, Category Partition offers a distinct trade-off. At the 20\% budget, it costs \$47 on SWE-bench Verified and \$31 on SWE-bench Multilingual, compared to \$45 and \$37 for Random. At the same time, its SccErr is lower on both benchmarks: 0.018 vs. 0.033 on Verified, and 0.033 vs. 0.042 on Multilingual. In these examples, selection improves the Success Rate estimate at a lower cost on Multilingual and a slightly higher cost on Verified.

\begin{figure*}[t]
\centering
\includegraphics[width= 0.8\textwidth]{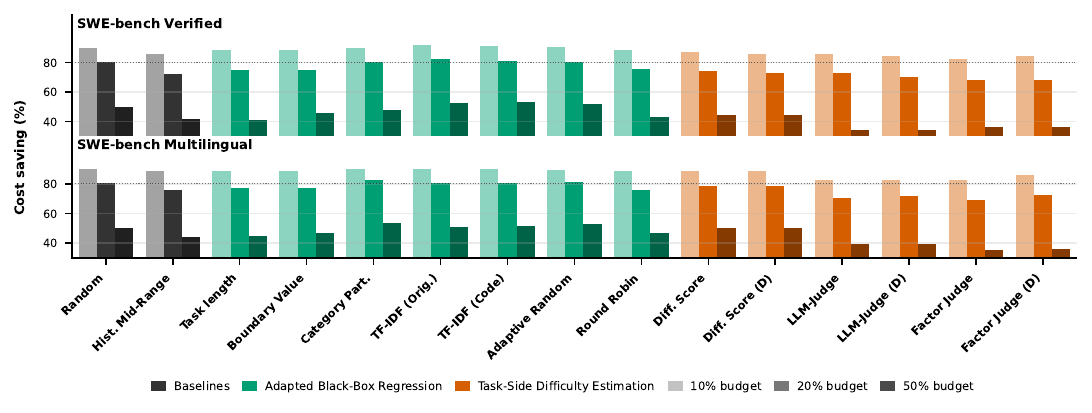}
\caption{\textcolor{black}{Cost saving (\%) at the 10\%, 20\%, and 50\% budgets (median over agents).}}
\label{fig:cost-saving}
\end{figure*}

\textcolor{black}{Also, figure~\ref{fig:cost-saving} shows the cost savings of each strategy at the 10\%, 20\%, and 50\% budgets. At the 20\% budget, several black-box strategies (Category Partition, TF-IDF, and Adaptive Random) save about 80-83\% of the cost. The judge-based strategies save less (about 68--73\%) because the tasks they select are more difficult and tend to be more expensive to run.}

\textbf{The cost of the LLM-as-judge}.
The strategies also differ in the cost of constructing their orderings. Random and most black-box strategies require no LLM calls, whereas LLM-judge strategies require difficulty estimates for the benchmark tasks prior to agent evaluation.

Table~\ref{tab:judge-cost} reports the cost of producing these judgments across the ten benchmarks. Using Claude Sonnet 5, the judgments cost approximately \$45.50 in total based on the recorded token usage. Using GPT-4o mini, the corresponding cost is approximately \$0.93. Importantly, each task is judged once, and the resulting difficulty scores can then be reused when evaluating different agents on the same benchmark. The cost of constructing the ordering can therefore be shared across multiple agent evaluations.
The utility of this additional step depends on the evaluation objective. For early failure detection, LLM-judge strategies provide clear improvements over Random and most black-box strategies, making the additional difficulty information valuable. For Success Rate estimation, judge-based strategies generally offer no corresponding advantage, as simpler black-box strategies already achieve low estimation error. The additional judging step is therefore most beneficial when the objective is to expose failures early.

\textbf{Judge model sensitivity: GPT-4o mini.}
The main experiments use Claude Sonnet 5 as the judge. To test sensitivity to this choice, we repeat the four LLM-judge strategies with GPT-4o mini, keeping the rest of the empirical methodology unchanged.
Figure~\ref{fig:cd-overall-f3} compares the difficulty-based strategies using the two judges. We omit the methods adapted from regression testing because they do not use judge scores and their results remain unchanged. For failure detection, the Task-Side Difficulty Estimation strategies rank above Random and Historical Mid-Range, while for Success Rate estimation, Random obtains the best mean rank. The connecting bars link strategies that are not statistically distinguishable under the Nemenyi test.
This analysis also reduces the concern that benchmark memorization drives the results. GPT-4o mini has a stated knowledge cutoff of October 2023, predating the release of several evaluated benchmarks, yet the overall pattern remains similar. This does not rule out possible training-data exposure for every benchmark.

\textbf{Implications for benchmark maintenance and agent development.}
The TCS methods can help curators examine benchmark composition and decide where to add new tasks. For example, Category Partition divides SWE-bench Verified into 89 partitions, of which 30 contain only one task, while the five largest contain 35.4\% of the benchmark. For Claude 4.5 Sonnet, Django tasks without traceback or exception names have a failure rate of 15.3\% in the shortest length group (9/59 tasks), compared with 50.0\% in the longest group (7/14); the full-benchmark failure rate is 28.6\%. Curators could add diverse tasks to sparsely populated or empty categories relevant to the benchmark's scope, or construct further tests for categories with high observed failure rates.

For agent developers, selecting tasks from a development or validation suite can reduce the cost of testing changes to the model, prompt, harness, tools, or execution environment. Inspecting the selected failures can help identify problems such as incorrect tool invocation or repository setup errors and guide harness improvements. On a held-out test set, selection can provide quick performance estimates and identify failures for inspection. 

\section{Threats to Validity}\label{sec:threats}

\textbf{Construct validity.}
We represent each task outcome as binary success or failure. For HAL results aggregated across repeated runs, a task counts as successful if the agent solves it in at least 50\% of runs. Fewer than 1.5\% of HAL task--agent outcomes are fractional, so this conversion affects few outcomes. For TerminalBench 2.0, more than 50\% of trials must pass. About one third of its task--agent results contain both successful and failed trials, so a different threshold could affect full-benchmark Success Rate and failure counts. We apply the same conversion rule to all strategies within each benchmark but do not evaluate alternative thresholds. Our difficulty-based strategies also assume that task information available before execution helps estimate difficulty. This assumption may not hold when difficulty depends mainly on tool behavior or execution conditions. Although Factor-Based LLM-as-Judge detects more failures than Random on all ten benchmarks at the 20\% budget (Table~\ref{tab:rq2}), these results may not generalize to other settings.

\textbf{Internal validity.}
Some strategies require benchmark-specific adaptations because the available task information differs across benchmarks. When a required field is unavailable, we use an alternative based on available task information, as described in Section~\ref{sec:methodology}, without using agent outcomes. Historical Mid-Range differs from the fourteen primary strategies because it uses other agents' outcomes, while excluding those of the target agent. LLM-as-judge strategies depend on the judge model and prompt. We repeat all four variants with GPT-4o mini and obtain a similar overall pattern, but do not systematically vary the prompts. Judges may have encountered benchmark tasks during training. GPT-4o mini's stated training cutoff of October 2023 predates several benchmark releases, reducing this concern but not ruling out the data leakage concern completely. We rely on published agent evaluation results and do not independently reproduce their outcomes. Errors in benchmark oracles, execution environments, or reported results could therefore affect our findings.




\textbf{External validity.}
We evaluate ten public agent benchmarks covering software engineering, competitive programming, terminal interaction, tool use, computational reproducibility, general assistant tasks, and web navigation. The results may not generalize to other domains, languages, or benchmarks. The benchmarks vary considerably in their numbers of tasks and evaluated agents. SWE-bench Mini overlaps with SWE-bench Verified and therefore does not provide an independent sample. The reported savings use recorded execution costs for the selected tasks; costs in subsequent evaluations may differ because agent behavior, API prices, and execution conditions can change.


\section{Conclusion and Future Work}

AI agents are increasingly used for software engineering, but evaluating them on complete benchmarks can be costly. To the best of our knowledge, this is the first empirical study of black-box test case selection for agentic software systems. We evaluated fourteen pre-execution selection strategies and a historical-performance baseline across ten public agent benchmarks. Our results suggest that strategies adapted from black-box regression testing are more effective for estimating an agent's full-benchmark Success Rate, while LLM-as-judge difficulty strategies detect more failures earlier. These overall patterns remained consistent across the different evaluated budgets.

Task selection can also substantially reduce evaluation costs. At the 20\% budget, Category Partition reduces the cost of evaluating GPT 5.2 Codex on SWE-bench Verified from \$225 to \$47 (a 79\% reduction) and DeepSeek V3.2 on SWE-bench Multilingual from \$114 to \$20 (an approximately 82\% reduction). The absolute error in Success Rate estimation is 0.8 percentage points in both cases. For early failure detection, Factor-Based LLM-as-Judge detects 53.0 failures for GPT 5.2 Codex on SWE-bench Verified at the 20\% budget, compared with 27.4 for Random, nearly twice as many within the same task budget. Overall, these results show that selecting benchmark tasks according to the evaluation objective can support performance estimation and failure detection at a fraction of the full execution cost.

Future studies can explore additional measures of task difficulty, such as cyclomatic complexity when task inputs include source code. Alternative embedding models, including general-purpose text encoders and other code models, can help assess how task representations affect selection. Beyond the TF-IDF methods evaluated here, other information-retrieval models could capture textual similarity~\cite{peng2020ir}. Alternatives to Round Robin could allocate selections proportionally to cluster size or maximize distance from already selected tasks. Finally, evaluating these approaches on new agents and benchmarks can further assess their generalizability.


\section*{Acknowledgments}
This work was conducted as part of the AI for Software Engineering (AI4SE) collaboration between JetBrains and Delft University of Technology. The authors gratefully acknowledge the financial support provided by JetBrains, which made this research possible.

\bibliographystyle{ACM-Reference-Format}
\bibliography{references}







\end{document}